\documentclass[aps,prb,reprint,twocolumn,groupedaddress,superscriptaddress,showpacs,floatfix,nobalancelastpage]{revtex4-1}
\usepackage{graphicx}
\usepackage{color}
\usepackage{hyperref}
\usepackage{amsmath}

\def \musr {$\mu^+$SR}

\begin{document}


\title{Spin dynamics in pressure-induced magnetically-ordered phases in PHCC}


\author{G.~Perren}
\affiliation{Neutron Scattering and Magnetism, Laboratory for Solid State Physics, ETH Z\"urich, CH-8093 Z\"urich, Switzerland}

\author{J.~S.~M\"{o}ller}
\affiliation{Neutron Scattering and Magnetism, Laboratory for Solid State Physics, ETH Z\"urich, CH-8093 Z\"urich, Switzerland}

\author{D. H\"uvonen}
\affiliation{Neutron Scattering and Magnetism, Laboratory for Solid State Physics, ETH Z\"urich, CH-8093 Z\"urich, Switzerland}
\affiliation{National Institute of Chemical Physics and Biophysics, 12618 Tallinn, Estonia}

\author{A.~A.~Podlesnyak}
\affiliation{Quantum Condensed Matter Division, Oak Ridge National Laboratory, Oak Ridge, Tennessee 37831-6473, USA}

\author{A.~Zheludev}
\homepage{http://www.neutron.ethz.ch/}
 \email{zhelud@ethz.ch}
\affiliation{Neutron Scattering and Magnetism, Laboratory for Solid State Physics, ETH Z\"urich, CH-8093 Z\"urich, Switzerland}


\date{\today}

\begin{abstract}
We present inelastic neutron scattering experiments on the $S=1/2$ frustrated gapped quantum magnet piperazinium hexachlorodicuprate under applied hydrostatic pressure. These results show that at 9~kbar the magnetic triplet excitations in the system are gapless, contrary to what was previously reported. We show that the changes in the excitation spectrum can be primarily attributed to the change in a single exchange pathway.
\end{abstract}

\pacs{75.10.Kt, 05.30.Rt, 62.50.-p, 78.70.Nx}


\maketitle

\section{Introduction}
Magnetic insulators offer a broad range of structural motifs, dimensionalities and well-defined short-ranged interactions. They are amenable to numerical modelling and are useful prototypes in the study of quantum phase transitions~\cite{Sachdev} (QPTs). An important and well-studied case of a QPT is the so-called Bose-Einstein condensation (BEC) of magnons in gapped quantum magnets in applied magnetic fields.\cite{Giamarchi2008} These are soft mode transitions, where at the QCP excitations have a parabolic dispersion, so the dynamical critical exponent $z=2$. A qualitatively different and much rarer soft mode transition can sometimes be induced in gapped spin systems through a continuous change of exchange constants. This, in turn, may in certain cases be achieved by the application of hydrostatic pressure. In these transitions the spectrum is expected to be linear at the quantum critical point (QCP), and hence $z=1$. Until recently only one good experimental realization of a pressure-induced QPT had been found, namely that in the three-dimensional dimer system TlCuCl$_3$.~\cite{Tanaka2003,Oosawa2004,Rueegg2004} Further work has lead to fascinating insights, in particular to the observation of a massive amplitude mode,~\cite{Rueegg2008TlCuCl3,Merchant2014} the magnetic analog of the Higgs boson.~\cite{SachdevKeimer}

Quantum magnets built from organic molecules can be very susceptible to perturbation by external pressure due to their `soft' molecular frameworks.~\cite{Saman2013} This paper is concerned with the $S=1/2$ quasi-two-dimensional gapped quantum antiferromagnet piperazinium hexachlorodicuprate [(C$_4$H$_{12}$N$_{2}$)Cu$_{2}$Cl$_{6}$, hereafter PHCC]. PHCC crystallizes in the triclinic space group $P\bar{1}$ with lattice parameters~\cite{Stone2001} $a=7.984(4)$~\AA, $b=7.054(4)$~\AA, $c=6.104(3)$~\AA, $\alpha=111.23(8)^\circ$, $\beta=99.95(9)^\circ$, $\gamma=81.26(7)^\circ$. The spin--$1/2$ Cu$^{2+}$ ions are connected by a complex layered network (see Fig.~\ref{fig:struc}). The magnetic interactions in PHCC have been studied using inelastic neutron scattering~\cite{Stone2001} (INS) and were found to be highly frustrated and more complicated than a simple dimer model. The ground state is a spin singlet separated by a gap $\Delta=0.98(6)$~meV from an $S=1$ triplet. More recent studies have demonstrated that the gap can be reduced by applied hydrostatic pressure, much as in TlCuCl$_3$.
At $p=9$~kbar the gap was found to decrease to $\Delta=0.55$~meV, and extrapolates to zero at $\sim 20$~kbar, hinting at a possible QPT at that point.~\cite{Hong2010PHCC}  In contradiction with this result, more recent muon-spin relaxation (\musr) experiments have discovered that the destruction of the spin-singlet state and the onset of magnetic long-range order occur at a much lower $p_{\rm c}\approx 4.3$~kbar.~\cite{Thede2014} A magnetically ordered phase of a Heisenberg spin system must have gapless spin waves, yet Ref.~\onlinecite{Hong2010PHCC} reported a spectral gap persisting well beyond the transition pressure observed in muon experiments.

The purpose of the present inelastic neutron study is to resolve the apparent controversy. Our new, higher resolution experiments reveal that the spin excitation spectrum in PHCC is gapless in the pressure-induced ordered state, already at 9~kbar. Moreover, we show that pressure-induced changes in the excitation spectrum can be primarily attributed to a variation in a single exchange pathway. The \musr\ experiments~\cite{Thede2014} found two distinct magnetically ordered phases: an incommensurate phase above a critical pressure $p_{\rm c}\approx 4.3$~kbar and a commensurate phase above $p_1\approx13.4$~kbar. Within the resolution of our experiment, we find no significant qualitative differences in the magnetic excitation spectra in the two regions.

\section{Experimental details}
\begin{figure*}[htb]
\includegraphics[width=\textwidth]{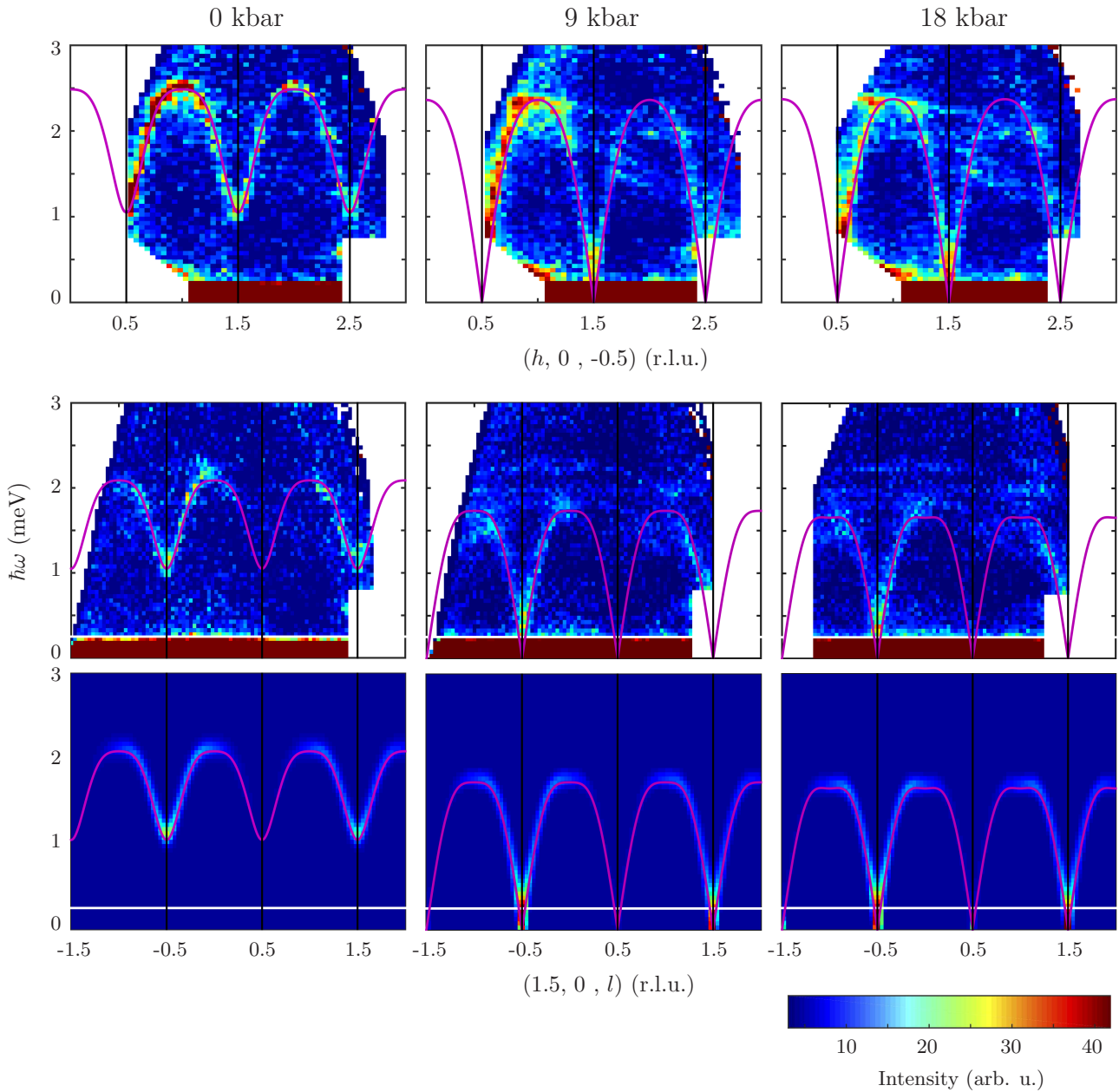}
\caption{\label{fig:3press}(color online). False-color slices through the 4-dimensional inelastic neutron data sets measured in PHCC at $T=1.5$~K, showing the dispersion of magnetic excitations in the $(h,0,l)$ plane.  The integration along the $k$ and $l$ or $h$ directions is $\pm 0.1$~r.l.u. Each column corresponds to a different applied pressure. Top and middle: background-subtracted experimental data. Bottom: simulation, based on the model cross section and the fitted parameter values, and convoluted with the resolution, as described in the text. In all cases, the solid lines are the model magnon dispersion relation, plotted using the fitted parameter values.}
\end{figure*}

High-quality deuterated single crystal samples of PHCC were grown by the thermal gradient method.~\cite{Yankova2012} The sample was a single crystal of mass 150~mg. Before this experiment, the sample was successfully used in a different neutron scattering study. The crystal was placed in a teflon tube and mounted with its $b$-direction vertical inside of a NiCrAl (``Russian alloy'') pressure cell. The top and bottom edges of the pressure cell (away from the sample) were covered in cadmium to reduce the background from the cell. Fluorinert was used as pressure-transmitting medium. The pressure was applied in a piston press, measured mechanically at ambient temperature and then extrapolated to low temperature using the documented behavior of the pressure cell. The pressure cell was mounted inside of a $^4$He cryostat on the time-of-flight cold neutron multi chopper spectrometer~\cite{Ehlers2011} (CNCS) at the Spallation Neutron Source (SNS), Oak Ridge National Laboratory. Data were collected with  incident energy of $E_{\rm i}=4.2$~meV at $T=1.5$~K. A converging guide was used to focus the neutron beam vertically onto the sample. The spectra were recorded by making 180$^\circ$ rotations with 1$^\circ$ step size. Corrections for the energy-dependent transmission of the pressure cell were performed using a reference measurement of incoherent elastic scattering from a plastic test sample.
Recorded events were projected onto the sample's reciprocal space coordinate system and binned into two-dimensional cuts with 150, 150, and 80 bins for $h$, $l$, and $\hbar \omega$, respectively, covering the range $(-3.7~{\rm r.l.u.}, -2.8~{\rm r.l.u.},  -0.5~{\rm meV})$ to $(3.7~{\rm r.l.u.}, 2.8~{\rm r.l.u.}, 3.5~{\rm meV})$. The data were integrated along $k$ by $\pm 0.1$ r.l.u.\ and exported for further analysis using the MANTID~\cite{Mantid} program.

\section{Results}
Our first result is that the lattice parameters $a$, $c$, and $\beta$ do not change abruptly between 0 and 18~kbar indicating the absence of a structural transition. From the position of the Bragg peaks in the scattering plane we found: $a=7.8(8) $~\AA, $c=6.0(6) $~\AA, $\beta=99.9(9)^{\circ}$ for 9 kbar and $a=7.8(8) $~\AA, $c=6.0(6) $~\AA, $\beta=100.2(2)^{\circ}$ for 18 kbar (both at 1.5 K). 

Selected inelastic spectra taken at 0, 9, and 18~kbar are shown in Fig.~\ref{fig:3press}. These are two-dimensional slices through the four-dimensional data set, integrated along the $k$ and $h$ (or $l$) directions in the range $\pm 0.1$~r.l.u. A smooth non-magnetic background was subtracted. For each cut, at each pressure, it was estimated by a second-order polynomial fitted to the intensity in areas where magnetic excitations are clearly absent. Even by eye it is immediately apparent that the excitations at 9~kbar are gapless with zero energy at the antiferromagnetic zone-center $(1.5,0,-0.5)$. This point coincides with the magnetic propagation vector of the ordered state that can be induced in PHCC at ambient pressure by an external magnetic field through a BEC-like transition.\cite{Stone2006PRL}

\begin{figure}[htbp]
\includegraphics[width=0.45\textwidth]{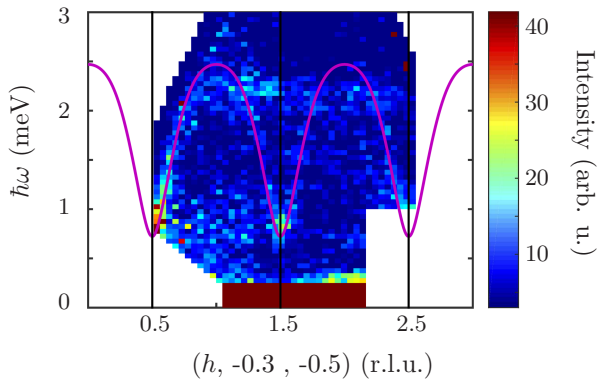}
\caption{\label{fig:kdep}(color online). False-color slice through the 4-dimensional inelastic neutron data sets measured in PHCC at $p=9$~kbar, $T=1.5$~K, showing the dispersion of magnetic excitations along the $(h,-0.3, -0.5)$ direction. The solid line is a guide for the eye.}
\end{figure}

Our setup, with the crystallographic $b$-axis mounted vertically, and with a vertically-focusing guide, is ill-suited for the study of the dispersion of excitations along the $b^\ast$ direction. Correspondingly, all the data shown in Fig.~\ref{fig:3press} are narrow slices around the $(h,0,l)$ plane. All the analysis given in the next section is applied only to the data taken in that plane. This is important to keep in mind, since the limited data available does indicate a small but significant dispersion along $b^\ast$. One such cut at 9~kbar, with $-0.4<k<-0.2$ is shown in Fig.~\ref{fig:kdep}. Note that at $h=1.5$ the excitation energy has increased to $\sim 0.7$~meV, as compared to $\sim 0$ for $k=0$  in the top-center panel of Fig.~\ref{fig:3press}.

\section{Analysis}
\begin{figure}[htb]
\includegraphics[width=\columnwidth]{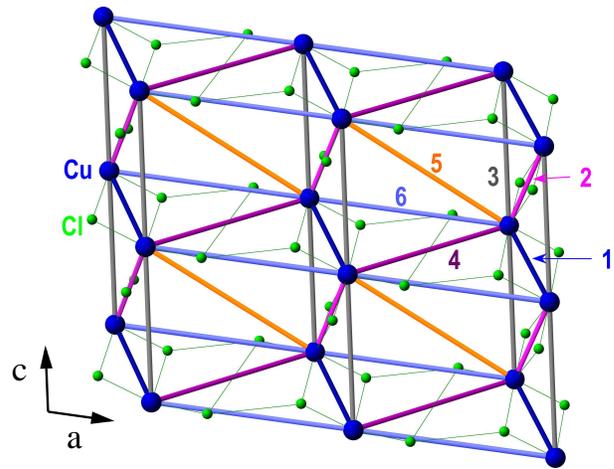}
\caption{\label{fig:struc}(color online). Structure of PHCC looking onto the crystallographic $(a,c)$ plane. Some bonds have been omitted for clarity. The relevant Cu-Cu super-exchange pathways are indexed for the discussion in the text.}
\end{figure}

Our analysis closely follows that used by Stone {\it et al.} in Ref.~\onlinecite{Stone2001}. To within experimental resolution, the scattering can in all cases be attributed to a single sharp magnon branch.
Correspondingly, the fitting model for the dynamic spin structure factor was written as:
\begin{equation}
S(\mathbf{Q},\omega) = S(\mathbf{Q}) {\delta(\hbar \omega - \hbar \omega_\mathbf{Q})}.
\label{eqn:dsf0}
\end{equation}
Note that we are not making any distinction between dynamic structure factors of different polarization, assuming spin correlations to be isotropic. Assuming PHCC is a Heisenberg spin system, this approach is fully justified at ambient pressure, where the ground state is disordered. However, in the magnetically ordered state the rotational symmetry is spontaneously broken. As a result, correlations between different spin components, as well as the corresponding contributions to the dynamic structure factor, will no longer be equal. Nevertheless, considering the resolution and noise level of our data, we applied the isotropic model  [Eq.~(\ref{eqn:dsf0})] as an empirical function at all experimental pressures.

In our model, also in line with Ref.~\onlinecite{Stone2001}, the equal-time structure factor $S(\mathbf{Q})$ was written as:
\begin{equation}
S(\mathbf{Q}) \propto \frac{1}{\hbar \omega_\mathbf{Q}} \sum_{\mathbf{d}} E_{\mathbf{d}}(\cos(\mathbf{Q}\cdot\mathbf{d})-1).
\label{eqn:dsf}
\end{equation}
In this expression the sum runs over all pairs of spins with relevant exchange interactions, separated by a bond $\mathbf{d}$. Once again, as long as the system is isotropic, this expression is well justified. It directly follows from the single-mode approximation [Eq.~(\ref{eqn:dsf0})], and the Hohenberg-Brinckman sum rule\cite{Hohenberg1974} for the first moment of the dynamic structure factor. As explained in detail for example in Ref.~\onlinecite{Zaliznyak2005b}, the coefficients $E_{\mathbf{d}}$ in this case are the expectation values of the Heisenberg exchange energy on the corresponding bonds between spins: $E_{\mathbf{d}} = J_{\mathbf{d}} \big \langle \mathbf{S}_0 \cdot \mathbf{S}_{\mathbf{d}} \big \rangle $. In the magnetically ordered (anisotropic) phase this interpretation is no longer valid. Nevertheless, in our analysis we used Eq.~\ref{eqn:dsf} as an empirical fitting function for data taken at all experimental pressures. In our case, the sum included all those bonds as in the approach of Ref.~\onlinecite{Stone2001}. The corresponding labeling scheme is shown in Fig.~\ref{fig:struc}.

The final ingredient of the model for $S(\mathbf{Q},\omega)$ is an empirical  dispersion relation which was parameterized at all pressures as:~\cite{Stone2001}
\begin{eqnarray}
(\hbar \omega_\mathbf{Q})^2 = & B_0 + B_h \cos(2\pi h) + B_l \cos(2\pi l) \\ \nonumber
&  + B_{hl} [\cos(2\pi(h+l)) + \cos(2\pi (h-l))]  \\ \nonumber
& + B_{2h} \cos(4\pi h) + B_{2l}\cos(4\pi l). \nonumber
\end{eqnarray}
Once again, we emphasize that our analysis was applied only to data in the $(h,0,l)$ plane. Correspondingly, the above expression does not include any dispersion along $b^\ast$.

For a direct comparison with the measured neutron intensities, the model structure factor given in Eq.~(\ref{eqn:dsf0}) was scaled with the magnetic form factor for Cu$^{2+}$ written in the dipole approximation, and numerically convoluted with the resolution function of our measurement. The latter was a direct product of Gaussian resolution functions for momentum transfer in the principle $h,l$ scattering plane and energy transfer, correspondingly.
The instrumental wave vector resolution was determined by a Gaussian fit to the width of the nuclear Bragg reflections in PHCC at very fine binning levels and was found to be practically isotropic (in the scattering plane) with $\sigma_Q^{\rm i}=0.054(2)$~\AA$^{-1}$. The instrumental energy resolution $\sigma_E^{\rm i}=0.069(3)$~meV was obtained by fitting the elastic line at a very fine binning level. 

For each pressure, the model was fitted simultaneously for the two two-dimensional data slices shown in Fig.~\ref{fig:3press}. At 9 and 18~kbar it was necessary to hold the parameter $B_{hl}$ fixed at the value fitted at ambient pressure. At 9 and 18~kbar, the upper part of the dispersion appears somewhat noisy and smeared out. The fits of the intensity (but not of the dispersion) therefore excluded the region above 2~meV to achieve reliable convergence. Table~\ref{tab:Bs} shows the resulting empirical fit parameters for the dispersion relation as well as the resulting spin gap and band widths derived from these values. Overall, our results at ambient pressure are in good agreement with the values obtained previously without the complication of performing an experiment inside of a pressure cell.~\cite{Stone2001} Our main result is that at 9~kbar, the gap $\Delta=0.0(1)$~meV, i.e.\ \emph{the spectrum is gapless}. Increasing the pressure to 18~kbar, the gap remains closed with $\Delta=0.0(1)$~meV.

Table~\ref{tab:Es} shows the fitted coefficients $E_{\mathbf{d}}$ at different pressures. Bonds 2--6 only exhibit modest changes in the whole 0--18~kbar region. The intensity coefficient for bond 1 however, changes dramatically between 0 and 9~kbar from $E_1=-1.11(5)$~meV at ambient pressure to $E_1=-0.47(3)$~meV at 9 kbar but changes only moderately to $E_1=-0.51(2)$~meV at 18~kbar.

\begin{table}[htb]
\begin{ruledtabular}
	\begin{tabular}{lrrrr}
		Parameter & Stone {\it et al.}\cite{Stone2001}    & 0~kbar & 9~kbar      & 18~kbar        \\ \hline
		$B_0$     & 5.44(2)  & 5.39(1)  & 4.29(1)  & 4.30(1)    \\ \hline
		$B _h$    & 2.06(3)  & 1.76(1) & 2.0(2)  & 2.0(2)    \\ \hline
		$B_l$     & 1.07(3)  & 0.85(1)  & 0.72(1)  & 0.59(1)    \\ \hline
		$B_{hl}$ & -0.39(1) & -0.39(5) & -0.39~~~ & -0.39~~~   \\ \hline
		$B_{2h}$ & -0.34(3) & -0.55(2) & -0.4(2) & -0.5(1)    \\ \hline
		$B_{2l}$ & -0.22(2) & -0.35(1) & -0.36(1) & -0.40(1)  \\ \hline \hline
		$\Delta$ & 0.98(6) & 1.05(3)  & 0.0(1) & 0.0(1)  \\ \hline
		$w_h$ & 1.60(8) & 1.44(4) & 2.4(1)  & 2.4(1)  \\ \hline
		$w_l$ & 1.18(8) &  1.04(4) & 1.7(1) & 1.7(1)  \\
	\end{tabular}
\end{ruledtabular}
\caption{\label{tab:Bs} Top: empirical dispersion parameters in meV$^2$ at different pressures. Bottom: resulting spin gap $\Delta$ and band width $w_h$ ($w_l$) along $h$ ($l$) in meV.}
\end{table}

\begin{table}[htb]
\begin{ruledtabular}
	\begin{tabular}{lrrrr}
		Bond energy & Stone {\it et al.}\cite{Stone2001}    & 0~kbar & 9~kbar     & 18~kbar       \\ \hline
		$E_1$ & -1.3(3)  &  -1.11(5) & -0.47(3) & -0.51(2)   \\ \hline
		$E_2$ & 0.7(3)   & 0.87(6)  & 0.77(4)  & 0.75(3)    \\ \hline
		$E_3$ & -0.3(1)  & -0.20(4)  & -0.14(3)   & -0.18(3)    \\ \hline
		$E_4$ & 0.1(3)   & 0.23(5)  & 0.22(3)    & 0.29(3)      \\ \hline
		$E_5$ & -0.0(3)  & 0.20(5)  & -0.08(3)   & -0.09(3)     \\ \hline
		$E_6$ & -0.92(5) & -0.85(4)  & -0.89(2) & -0.89(2)
	\end{tabular}
\end{ruledtabular}
\caption{\label{tab:Es} Fitted intensity modulation coefficients $E_{\mathbf{d}}$ in meV, for different pressures. The bond labeling corresponds to that in Fig.~\ref{fig:struc}.}
\end{table}

\section{Discussion}
Our results demonstrate unambiguously that the magnetic excitations in PHCC are gapless at and above 9~kbar applied pressure. This is fully in agreement with the observation of long-range magnetic order above a QCP at $p_{\rm c} \approx 4.3$~kbar using \musr.~\cite{Thede2014} A pressure-induced reduction of the gap is also plausible in the context of the observation that Br-substitution, which leads to an increase of the lattice parameters and hence \emph{negative} chemical pressure, causes the gap to \emph{increase}.\cite{Huevonen2013} The discrepancy with Ref.~\onlinecite{Hong2010PHCC}, where the gap seemed to extrapolate to zero only at much larger pressures, warrants an explanation.
In that study, PHCC was assumed to possess a two-dimensional magnon dispersion. Under this assumption, to improve statistics, the data were integrated along the $k$-direction in a substantial range. As discussed above, the assumption is clearly invalid. Due to the dispersion along $b^\ast$, any integration will produce an intensity maximum at the saddle-point energy (around 0.7~meV). We suggest that it is this saddle-point that was mistaken for the energy gap in Ref.~\onlinecite{Hong2010PHCC}.

As stated above, for $p>p_c$, in the ordered state, the parameters $E_\mathbf{d}$ cannot be directly interpreted as bond energies due to the anisotropic nature of the ordered state. This said, the fact that our isotropic model gives reasonably good fits to the experimental data in all cases, suggests that the anisotropy of spin correlations at high pressures remains small. This, in turn, suggests that the ordered moment remains small compared to the classical saturation value. In this case, a consistent and very pronounced decrease of $|E_1|$ with increasing pressure, and a lack of any drastic pressure dependence for the other parameters in Table~\ref{fig:struc}, are evidence that the transition is, in fact, driven by the weakening of a single bond. That bond, labeled as ``1'' in Fig.~\ref{fig:struc} and in Ref.~\onlinecite{Stone2001}, happens to have the largest bond energy at ambient pressure. Furthermore, we note that we have not been able to observe a magnetic Bragg peak in the magnetically ordered phases which would be consistent with a small ordered moment size.

We recall that there is some experimental \musr\ evidence that the magnetically-ordered state emerging at $p_c$ is incommensurate.~\cite{Thede2014} Our wave vector resolution is approximately 0.07, 0.2 and 0.05 reciprocal lattice units FWHM along $h$, $k$, and $l$, respectively. This sets an upper bound on the magnitude of the incommensurate propagation vector in the system. Beyond that, the present study is unable to confirm or disprove the incommensurate nature of the pressure-induced magnetic ordering. As far as the fitted dispersion and bond coefficients $E_{\mathbf{d}}$ are concerned, only moderate changes are observed between 9~kbar (in the first phase detected with  \musr) and 18~kbar (in the second phase found with \musr).

\section{Conclusion}
In summary, it appears that the magnetically ordered state induced in PHCC by hydrostatic pressure in excess of $p_c\sim 4.3$~kbar is gapless. The transition itself is driven primarily by the weakening of one particular exchange pathway. At high pressures PHCC is by no means a two-dimensional system, with a substantial magnon dispersion along the third ($b^\ast$) direction.

\section*{Acknowledgements}
We thank Georg Ehlers for technical assistance and David Schmidiger for useful discussions. This work is partially supported by the Swiss National Fund. Research at Oak Ridge National Laboratory's Spallation Neutron Source was supported by the Scientific User Facilities Division, Office of Basic Energy Sciences, US Department of Energy. JSM gratefully acknowledges support through an ETH Fellowship. DH acknowledges support by the Estonian Ministry of Education and Research under grant IUT23-03 and Estonian Research Council grant PUT451.


\bibliography{phcc}

\end{document}